\documentclass[%
 reprint,
 amsmath,amssymb,
 aps,
]{revtex4-2}

\usepackage{graphicx}% Include figure files
\usepackage{dcolumn}% Align table columns on decimal point
\usepackage{bm}% bold math
\usepackage{xcolor}

\usepackage{tikz}
\usetikzlibrary{arrows.meta}

\definecolor{plum}{rgb}{0.36078, 0.20784, 0.4}
\definecolor{chameleon}{rgb}{0.30588, 0.60392, 0.023529}
\definecolor{cornflower}{rgb}{0.12549, 0.29020, 0.52941}
\definecolor{scarlet}{rgb}{0.8, 0, 0}
\definecolor{brick}{rgb}{0.64314, 0, 0}
\definecolor{sunrise}{rgb}{0.80784, 0.36078, 0}
\definecolor{lightblue}{rgb}{0.15,0.35,0.75}
\definecolor{carolina}{RGB}{153, 186, 221}
\tikzstyle{axisarrow} = [-{Latex[inset=0pt,length=5pt]}]

\begin{document}

\preprint{APS/123-QED}

\title{A unit of information in black hole evaporation}
%correspondence to Rindler-Unruh spacetime} %$: cosmic no-hair?} %: $\mathbb{P}^2$ and $S^2$}

\author{Maurice H.P.M. van Putten}
\altaffiliation[]{{ INAF-OAS Bologna, via P. Gobetti, 101, I-40129 Bologna, Italy and Department of} Physics and Astronomy, Sejong University}
 \email{maurice.vanputten@inaf.it}
%\author{author2}%
%\altaffiliation[]{author2 affiliation}

\date{\today}

\begin{abstract}
{Black holes evolve by evaporation of their event horizon. 
While this process is believed to be unitary, there is no consensus on the recovery of information in black hole entropy. 
A missing link is a unit of information in black hole evaporation. 
Distinct from Hawking radiation, 
we identify evaporation in entangled pairs by $\mathbb{P}^2$ topology of the event horizon consistent with the Bekenstein-Hawking 
entropy in a uniformly spaced horizon area, where $k_B$ denotes the Boltzmann constant. 
It  derives by continuation of $\mathbb{P}^2$ in Rindler spacetime prior to gravitational collapse, subject to a tight correlation of the
fundamental frequency of Quasi-Normal-Mode (QNM) ringing in gravitational and electromagnetic radiation.
Information extraction from entangled pairs by detecting one over the surface spanned by three faces of a large cube carries a unit of information of $2\log3$
upon including measurement of spin.}
%We comment on a potential observational consequences for the fundamental frequency of Quasi-Normal-Mode ringing.} 
\end{abstract}


\maketitle

\section{Introduction}

%\noindent {\bf\large Introduction}\\
%\noindent 
%{\em Introduction.} 
Modern observations reveal a Universe rife with supermassive and stellar mass black holes  
\citep{gen03,ghe03,ghe08,gen10,abb16,EHT22} in a cosmological background showing accelerated 
expansion on a Hubble scale \citep{agh20,rie21}.

Governed by gravitation across these astrophysical and cosmological scales, event horizons define a common theme regardless of spacetime curvature.
Presciently introduced in cosmological spacetime \citep{rin60}, a principle example is the Rindler horizon to accelerating observers.
Independent of general relativity, it is recognized to have a finite temperature \citep{ful73,dav75,unr76,tho84,tho85}.
Thermodynamics of black holes and de Sitter spacetime \citep{haw74,haw75,gib77} is hereby anticipated by that of Rindler spacetime.

The Bekenstein-Hawking entropy \citep{bek73,haw75} in surface area of black hole event horizons $H$ reveals a first step beyond the no-hair theorem of general relativity and no-memory of black hole formation history \citep{pen65}. 
In a unitary evolution, its formation history is expect to be recovered by information in Hawking radiation, which continues to elude the unification of gravitation and quantum field theory \citep[e.g.][]{cal22}.

A one-to-one equivalence to other horizon surfaces need not hold, however, possibly due to non-local properties of the spacetime vacuum.
While black holes are defined by mass and angular momentum at asymptotically flat Minkowski spacetime ${\cal M}$, the same is prohibitive in cosmology in the face of a cosmological horizon at a finite Hubble distance.
Evolution in cosmology is governed by content in dimensionless densities, normalized to closure density $\rho_c = 3H^2/8\pi G$ given Newton's constant $G$.

While {\em Planck} analysis of the CMB \citep{agh20} strongly supports $\Lambda$CDM, primarily parameterized by content in dark energy and dark and baryonic matter, this is challenged by model-independent observations in the Local Distance Ladder \citep{rie21}. These observations may indicate potential anomalies in electromagnetic propagation and our conventional Friedmann-Lema\^itre-Robertson-Walker (FLRW) cosmologies \citep[e.g.][]{csa02a,csa02b,dam17,col19,spa21,spa21,spa22,sar22}. 
Alternatively, cosmological spacetime is endowed with properties beyond the classical limit described by general relativity, e.g., a dark energy inherent to quantum cosmology \citep{van21}.

A missing link in black hole evaporation is a unit of information in the detection its radiation at large distances.
A discrete unit is essential to any exact formulation of information encoding in unitary evolution \citep{van15a}.
We here approach this by horizon topology distinct from classical no-hair - the trivial topology of a point with finite surface area. 
This second step beyond general relativity is found to be in exact correspondence with a uniform area spacing of black hole event horizons \citep{hod98}, further consistent with gravitational radiation \citep{abb16}.
 
To this end, we consider the horizon topology by continuation of null-generators of $H$ back in time to the light cone in ${\cal M}$ prior 
to gravitational collapse, probed in the Rindler spacetime of test particles suspended by a distant observer (Fig. \ref{FIGR}).
By the equivalence principle, this probe nearby and along these null-generators is constant to a distant observer, 
traces the topology of $H$ back to that of the latter in Rindler spacetime ${\cal R}$.

With no appeal to general relativity, our approach is distinct from earlier studies of the Kruskal extension of eternal black holes as classical solutions to the Einstein equations \citep{tho85,gib86,cha97,lou97,tho21}.
Whether results thereof apply to black holes formed in gravitational collapse is subject to ambiguities, related to double cover spaces \citep{gib86,tho21} with associated (non-)orientability in their embedding in higher dimensions \citep[cf.][]{cha97}. 

In \S2, we revisit the definition of Rindler spacetime and show it to produce the Schwarzschild radius of a black hole formed in gravitational collapse.
In \S3, we discuss the consequences for temperature and topology of Schwarzschild black holes and compare the results with the cosmological horizon of de Sitter space. 
In \S4, an ambiguity in horizon temperature is discussed very similar to those mentioned above following the original proposal of a potential double Hawking temperature $2T_H$ \citep{tho84,tho85}. Preserving the correlation of the fundamental frequency of Quasi-Normal-Mode (QNM) ringing and Hawking temperature points towards the decay of $2k_BT_H$ energy transitions to entangled pairs at the Hawking temperature $T_H$, rather than radiation of Hawking particles of energy $2k_BT_H$. We summarize our findings in \S5.

\section{Rindler spacetime revisited}

Observers at constant acceleration $a$ appear along hyperbolic world-lines in Minkowski spacetime ${\cal M}$ of inertial observers in 1+1 spacetime $(t,x)$. 
These world-lines appear straight in Rindler spacetime ${\cal R}$ in Rindler coordinates $(\lambda,\xi)$. In the conventional definition, ${\cal R}$ covers a single wedge of ${\cal M}$. 
Following \cite{rin60}, the principle invariant in Rindler spacetime is the velocity of light $c$, satisfying
\begin{eqnarray}
c=\sqrt{a\xi},
\label{EQN_c}
\end{eqnarray}
where $\xi$ denotes the Lorentz invariant distance to the Rindler horizon $h$.
It implies some elementary equivalences:
%in ${\cal R}$ with gravitational field $g=-a$ by the equivalence principle:
(i) the redshift factor $\delta$ of a photon source at $0\le \delta \xi \le \xi$ seen in ${\cal R}$ has an equivalent to Doppler shift to red by the Lorentz factor $\Gamma$ seen in ${\cal M}$;
(ii) inertia defined by mass-energy in binding energy $U_h = \int_0^\xi mads$ to $h$ equals to $mc^2$ \citep{n4}, provided $h$ is within the Hubble horizon \citep{van17}. 
 
Rindler spacetime ${\cal R}$ can be viewed as the conformal image of ${\cal M}(t,x))$ in $z=(it,x)\,\epsilon\,\mathbb{C}$ by
\begin{eqnarray}
w(z)=\xi^{-1}z^2.
\label{EQN_w}
\end{eqnarray}
Hyperbolic world-lines $z=\xi\left(\cosh\lambda+i\sinh\lambda\right)$ with $\Gamma=\cosh\lambda$ in ${\cal M}$ are parameterized by a
rapidity $\lambda=a\tau/c^2$ at eigentime $\tau$ and invariant arclength $ds=\xi d\lambda$ at distance $\xi=c^2/a$ from the origin.
By (\ref{EQN_w}), the image is a straight line $w=\xi\left(1+i\sinh(2\lambda)\right)$ in ${\cal R}$ conform the line-element $ds^2=dx^2-c^2dt^2=d\xi^2 - c^2d\tau^2 = d\xi^2 + \xi^2d\theta^2$ in polar coordinates $(\xi,\theta)=(\xi,i\lambda)$.
The period $\pi i$ of $\lambda$ in the trajectory in ${\cal R}$ is one-half the period $2\pi i$ of the hyperbolic trajectory in ${\cal M}$ \citep{n3}. 
By (\ref{EQN_w}), the left of $h$ is the image of the future and past $(\left|t\right|>\left|x\right|)$ of the origin in ${\cal M}$.  

Fig. 1 shows a probe of gravitational collapse by a distant observer, measuring the tension in a long rope suspending a test particle $m$. 
This probe measures the gravitational field close to a null-geodesic, initially along a light cone in ${\cal M}$ that evolves into a generator of the event horizon around $M$ after collapse. Tension in the rope is initially due to inertia in ${\cal R}$ and, at late times, dues to gravitational attraction to $M$. As these two alternatives are indistinguishable by the equivalence principle, the distant observer may attribute both to inertia in ${\cal R}$. 

\begin{figure*}
%    \centering
    \includegraphics[scale=0.24]{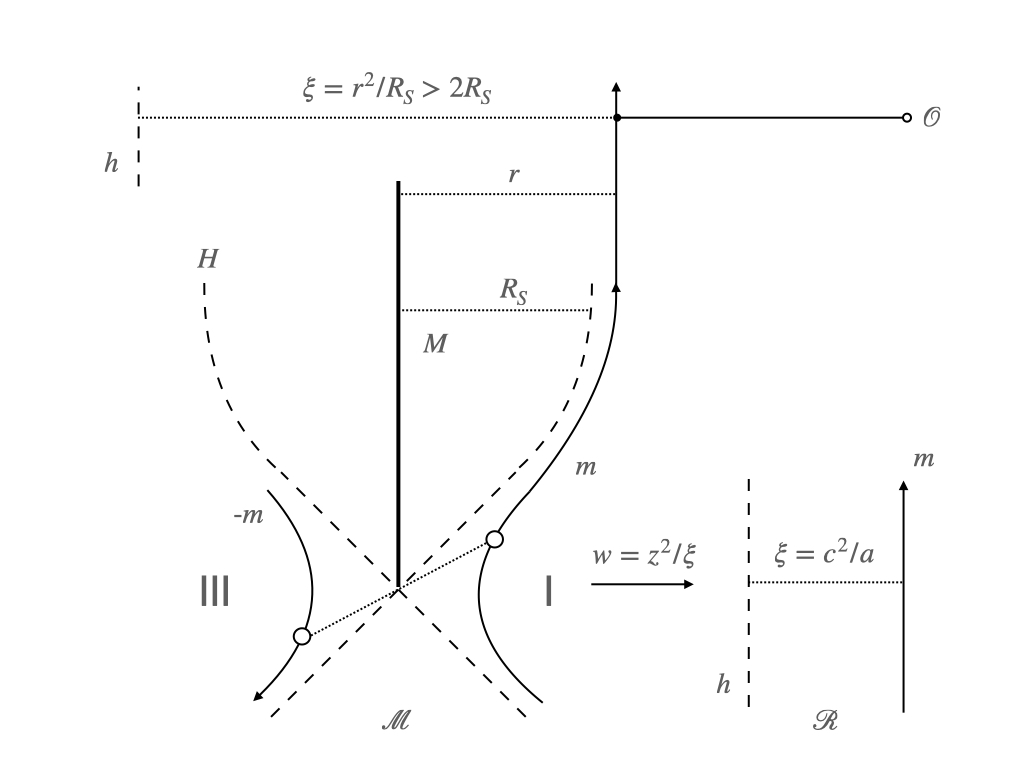}
    \includegraphics[scale=0.24]{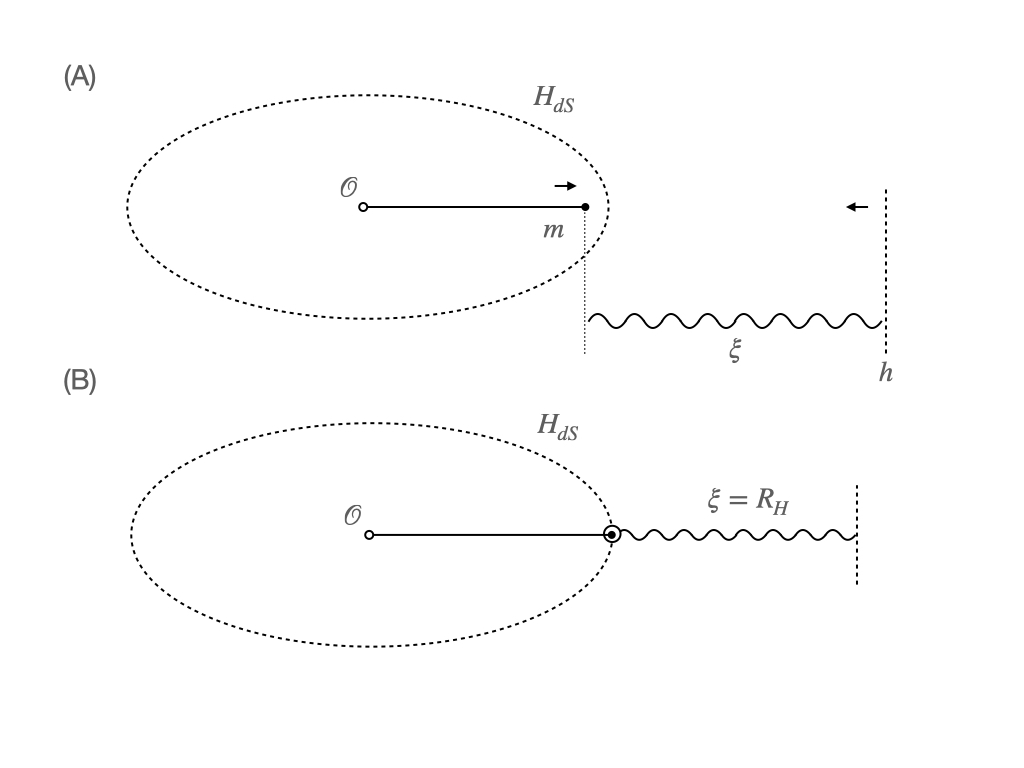}
    \vskip-0.0in
    \caption{
    (Left panels.) 
    Prior to gravitational collapse, a distant observer infers $\mathbb{P}^2$ of the light cone in ${\cal M}$ by its double cover (wedges I and III in ${\cal M}$) of the Rindler space of a test particle $m$, held in suspension along a hyperbolic trajectory close to a null-generator of this light cone.
    $\mathbb{P}^2$ topology of $H$ about $M$ follows by continuation of
    (dashed lines) during gravitational collapse and the equivalence principle.
    By spherical symmetry of the observer's position on the celestial sphere, $H$ 
    is a sphere of infinite redshift inferred in the limit as the Rindler horizon $h$ of $m$ and $m$ approach $M$ at equal distance, satisfying $r=R_S$. 
    (Right panels.) A similar procedure applied to ${\cal R}$ of ${\cal O}$ suspending $m$ above the cosmological horizon $H_{dS}$
    of de Sitter (A). As $m$ approaches $H_{dS}$, $h$ extends beyond $H_{dS}$ at a Hubble distance $R_H$ (B). $H_{dS}$ herein assumes topology $S^2$.
    }
    \label{FIGR}
\end{figure*}

In this probe, the observer infers an event horizon with the Schwarzchild radius of $M$ with no appeal to general relativity.
Given the gravitational radius $R_g=GM/c^2$ and post-collapse, the observer suspends $m$ suspended at rest at a distance $r$ above $M$. The observer infers a Rindler horizon $h$ to $m$ at distance
\begin{eqnarray}
 \xi = \frac{r^2}{R_g}
\label{EQN_xi}
\end{eqnarray}
by virtue of Newton's law $a=GM/r^2$. Thus, $\xi \ge 2r$ (Fig. \ref{FIGR}).
By spherical symmetry in the position of the observer on the celestial sphere, 
$\xi = 2r$ is a limiting case, when $m$ assumes the infinite redshift of $h$.
$M$ hereby possesses a spherical surface of infinite redshift $H$ 
at $r=R_S$ equal to the Schwarzschild radius, 
\begin{eqnarray}
    R_S=2R_g.
    \label{EQN_RS}
\end{eqnarray}

Entanglement entropy $\Delta S = 2\pi \Delta \varphi$ of a mass-perturbation $dM$ relative to $h$ in ${\cal R}$ \citep{n4} of a Compton phase $d\varphi = \xi dR_g /l_p^2$ across $\xi^\prime =2R_S=4R_g$ \citep{n4} can be integrated to 
$k_B^{-1}dS=2\pi d\varphi$ over $M$. Here, $k_B$ is the Boltzmann constant and 
$l_p^2=G\hbar/c^3$ is the Planck area given Planck's constant $\hbar$.
In units of $l_p^2$, this recovers the Bekenstein-Hawking entropy
\begin{eqnarray}
S=2\pi k_B \int_0^M \xi dR_g = 4\pi k_B R_g^2 = \frac{1}{4}k_B A_H.
\label{EQN_SH}
\end{eqnarray}
Crucially, (\ref{EQN_RS}-\ref{EQN_SH}) derive strictly from (flat) ${\cal R}$ 
with no appeal to curved spacetime of general relativity. 

\section{Temperature and topology}

In Fig. 1, universality in the equivalence principle implies a constant vacuum temperature measured according to acceleration prior to collapse and surface gravity post-collapse. That is, the temperature of $H$ about $M$ is that of ${\cal R}$ of the test mass inferred by the distant observer.

For a Davies-Unruh temperature of $h$ \citep{ful73,dav75,unr76}, Newton's law $a=GM/r^2$ in the above implies a horizon temperature 
\begin{eqnarray}
 k_BT_H = \frac{\hbar c}{8\pi R_g}
\label{EQN_H}
\end{eqnarray}
at $r=R_S$, i.e., the Hawking temperature of a black hole. Combined with
(\ref{EQN_SH}), (\ref{EQN_H}) identifies the mass-energy $M c^2$ with heat in the Clausius integral $Q=\int_0^M T_HdS$.

The result (\ref{EQN_H}) may be compared with the temperature $T_{dS}$ of de Sitter vacua according to the surface gravity of the Sitter horizon $H_{dS}$ 
at the Hubble radius $R_H$. 
Suspended by a rope at a distance $r$, a test particle $m$ experiences non-geodesic motion at a velocity $v=Hr$ relative to the local Hubble flow.
Approaching $H_{dS}$ in the limit of $v=c$, $m$ carries $h$ at distance 
\begin{eqnarray}
 \xi \ge R_H
\end{eqnarray}
according to (\ref{EQN_c}), $a=rH^2$, and hence
\begin{eqnarray}
    \xi = \frac{R_H^2}{r}
    \label{EQN_xidS}
\end{eqnarray}
within the visible universe $r\le R_H$. 
Surface gravity of $H_{dS}$ is the de Sitter scale of acceleration $a_{dS}=cH$ \citep{van17}. 
For a Davies-Unruh temperature to $h$ once more, ${\cal O}$ identifies a de Sitter temperature \citep{gib77}
\begin{eqnarray}
    k_BT_{dS} = \frac{\hbar c}{4\pi R_g}.
\label{EQN_dS}
\end{eqnarray}
Here, $R_g=R_H/2$ denotes the gravitational radius of a total dark energy 
content at the limit of dimensionless dark energy density $\Omega_\Lambda=1$, 
i.e., $\rho_\Lambda$ in de Sitter heat at closure density satisfying the anomalous scaling $\rho_c\propto T_{dS}^2$.

By (\ref{EQN_w}), ${\cal R}$ is double covered by the two spacelike wedges in ${\cal M}$. To a distant observer, the light cone in ${\cal M}$ hereby assumes $\mathbb{P}^2$ topology. $H$ inherits the same by the equivalence principle in the continuation of generators following gravitational collapse.

\section{Hawking particles or entangled pairs?}

In Fig. 1, we identify ${S}^2$ of the de Sitter horizon surfaces $H_{dS}$.
Given $R_g$, this discrepant topology may account for (\ref{EQN_H}) being lower than (\ref{EQN_dS}) by a factor of two.
It would represent universal scaling of horizon temperature with surface gravity in \ref{EQN_H}) and (\ref{EQN_dS}).

By (\ref{EQN_w}), ${\cal R}$ is doubly covered by the wedges I and III of ${\cal M}$.
The resulting $\mathbb{P}^2$ topology in $H$ is distinct from $S^2$ that would derive from a conventional interpretation of Rindler spacetime, covered by a single wedge I in traditional Rindler coordinates \citep[e.g.][]{val23}. Accordingly, it assumes the conventional Davies-Unruh temperature.

By (\ref{EQN_w}), the imaginary period $i\pi$ in the rapidity $\lambda$ of the world-line is one-half the period $2\pi i$ of its hyperbolic pre-image in ${\cal M}$.
By this change in periodicity in the Killing direction $\theta=i\lambda$ in $ds^2=d\xi^2+\xi^2d\theta^2$, the temperature of ${\cal R}$ under (\ref{EQN_w}) is hereby twice the Davies-Unruh temperature. By continuation in Fig. 1, this implies a doubling of Hawking temperature by continuation. Independently, the same conclusion has been derived from the Kruskal extension of Schwarzschild black holes by 't Hooft \citep{tho84,tho85,tho21}. 

By $\mathbb{P}^2$, a temperature $2T_H$ of in (\ref{EQN_H}) now satisfies the same scaling in gravitational radius as the temperature $T_{dS}$ of de Sitter horizon  (\ref{EQN_dS}), different from universal scaling with surface gravity. 
Accompanying this increase in temperature is a reduction in entropy conform $S=\hbar \omega/k_BT$ for a photon at energy $\hbar\omega$ \citep{n4}, reducing black hole entropy (\ref{EQN_SH}) by a factor of two.

However, the Hawking temperature scales directly with the fundamental frequency of Quasi-Normal-Mode (QNM) ringing of the event horizon of a Schwarzschild black hole \citep{cha75,sch85,lea85,lea86,kok99,ma23}, given \citep{hod98}
\begin{eqnarray}
 {\sc Re}\, \hbar\omega \simeq \log 3\, k_BT_H
 \label{EQN_QNM}
\end{eqnarray}
in the large-$n$ limit of overtones $n$. It derives from linear perturbation theory of black holes in classical general relativity \citep[e.g.][]{hod98}.
This may be attributed to the fact that the fundamental frequency of horizon modes is independent of spin of the radiation field. 
While a detailed consideration of the correlation (\ref{EQN_QNM}) in $\mathbb{P}^2$ falls outside the scope of the present work, such is expected to persist to preserve (\ref{EQN_i}). 
A doubling of $T_H$ would hereby be indicated by a doubling of the fundamental frequency $\omega$ of QNM. {\em This seems unlikely.} 
In fact, it will be tested by black hole spectroscopy \citep{ech89,dre04,ber06,ber07,car16,ber09} in the LIGO-Virgo-KAGRA (LVK) observational runs O4-5.

Instead, the tight correlation (\ref{EQN_QNM}) is preserved in the energy spectrum of black hole decay at double the Hawking temperature
into entangled pairs rather than Hawking particles in conventional black body radiation.
{That is, entangled pairs preserve the Hawking temperature in black hole decay at double the Hawking temperature.}

\section{Conclusions and outlook} 

With no appeal to general relativity of curved spacetime, we infer $\mathbb{P}^2$ topology of $H$ with Schwarzschild radius $R_S=2R_g$ by equivalence to Rindler spacetime, by continuation along null-geodesics through the process of gravitational collapse. 
This topology is beyond the no-hair theorem of black holes in general relatiivity. 

Preserving (\ref{EQN_QNM}),  $\mathbb{P}^2$ points to black hole evaporation in entangled pairs at the Hawking temperature 
rather than Hawking particles at twice the Hawking temperature.
While information encoding occurs at emission, information retrieval is defined by measurement outcome at detection. 
The first occurs unseen over a continuum in amplitude of directions subject to $\mathbb{P}^2$ in emission from the horizon.
In retrieving information from entangled pairs, it suffices to detect one of the two. This can be realized over three states in detection
over the surface spanned by three faces of a large cube normal to $+x$, $+y$ and $+z$ in a given Cartesian coordinate system $(x,y,z)$ centered about the black hole. 
Thus, each detection retrieves
\begin{eqnarray}
   i = 2\log 3
   \label{EQN_i}
\end{eqnarray}
of information upon including information in spin $\left\{-1,0,1\right\}$, e.g., with reference to the directions $(+x,+y,+z)$. 
Crucially, (\ref{EQN_i}) is {\em independent of the energy spectrum}. 

Distinct from conventional the unit $k_B\log2$ in binary encoding, (\ref{EQN_i}) equals twice the unit of Bekenstein-Hawking entropy, 
defined by the uniform area spacing $4 \log 3$ in units of Planck area $l_p^2$ \citep{hod98}.
This correspondence appears relevant, taking into account evaporation in emission one-by-one \citep{bek95,van15b}, here of entangled pairs with 
mean rate of detecting  (\ref{EQN_i}) at $\dot{n}_\gamma \simeq 40\,\left({M}/{M_\odot}\right)^{-1}\,{\rm Hz}$ \citep{spa16}.

$\mathbb{P}^2$ hereby sets a starting point for unitary evolution of black holes by evaporation, expected to retrieve all of the Bekenstein-Hawking entropy of classical no-hair \citep[e.g.][]{tho21,cal22}. 
A $\mathbb{P}^2$ topology of $H$ is also natural, consistent with spin-2 of gravitational radiation \citep[e.g.][]{kok99}, 
now observed by direct detection from binary black hole mergers \citep{abb16} and from the central engine of the short GRB170817A \citep{van23}. 

%Bipolar outflows: astrophysical jets 

\begin{acknowledgments}
%The author thanks the anonymous reviewer 
We gratefully acknowledge G. 't Hooft for detailed discussions on a possible ambiguity in the temperature of black holes and discussions
with C. Nunez and D. Thompson over a theory seminar at Swansea University, UK (2022).
\end{acknowledgments}


\begin{thebibliography}{99}
\bibitem[Genzel et al.(2003)]{gen03} Genzel, R., Sch\"odel, R., Ott, T., et al., 2003, ApJ, 594, 812
\bibitem[Genzel et al.(2010)]{gen10} Genzel, R., Eisenhauer, F., \& Gillessen, S., 2010, RMP, 82, 3121
\bibitem[Ghez et al.(2003)]{ghe03} Ghez, A.M., Duch\^ene, G.,Matthews, K.,  et al., 2003, ApJ, 586, L127
\bibitem[Ghez et al.(2008)]{ghe08} Ghez, A., Salim, S., Weinberg, N.N., et al., 2008, ApJ, 689, 1044
\bibitem[EHT(2022)]{EHT22} EHT Collaboration: Akiyama, K., Alberdi, A., Alef, W., et al., 2022 ApJL 930 L17
\bibitem[Abbott et al.(2016)]{abb16} LIGO-Virgo Collaboration: Abbott, B.P., Abbott, R., Abbott, T.D., 2016, et al. Phys. Rev. Lett., 116, 061102

\bibitem[Aghanim(2020)]{agh20} Planck Collaboration: Aghanim, N., Akrami, Y., Ashdown, M., et al. 2020, A\&A, 641, A6
\bibitem[Riess et al.(2021)]{rie21} Riess, A.G., Yuan, W., Macri, L.M., et al., 2022, ApJ, 934, L7 
\bibitem[Rindler(1960)]{rin60} Rindler, W., 1960, Phys. Rev., 119, 2082

\bibitem[Fulling(1973)]{ful73} Fulling, S.A., 1973, Phys. Rev. D, 7, 2850
\bibitem[Davies(1975)]{dav75} Davies, P.C.W., 1975, J. Phys. A: Math. Gen., 8, 609
\bibitem[Unruh(1976)]{unr76} Unruh, W.G., 1976, Phys. Rev. D, 14, 870 

\bibitem['t Hooft(1984)]{tho84} 't Hooft, G., 1984, J. Geom. Phys, 1, 45
\bibitem['t Hooft(1985)]{tho85} 't Hooft, G., 1985, Nucl. Phys. B, 256, 727

\bibitem[Hawking(1974)]{haw74} Hawking, S., 1974, Nat., 248, 30
\bibitem[Hawking(1975)]{haw75} Hawking, S., 1975, Commun. Math. Phys. 43, 199
\bibitem[Gibbons\&Hawking(1977)]{gib77} Gibbons, G.W., \& Hawking, S.W., 1977, Phys. Rev. D, 15, 2738

\bibitem[Bekenstein(1973)]{bek73} Bekenstein, J.S., 1973, 1073, Phys. Rev. D 7, 2333
\bibitem[Penrose(1965)]{pen65} Penrose, R., 1965, Phys. Rev. Lett., 14, 57

\bibitem[Calmet \& Hsu(2022)]{cal22} Calmet, X., \& Hsu, S.D.H., 2022, Eur. Phys. Lett., 139, 49001

\bibitem[Cs\'a et al.(2002)]{csa02a} Cs\'a, C., Kaloper, N., \& Terning, J., 2002, Phys. Rev. Lett., 88, 161302
\bibitem[Cs\'a et al.(2002)]{csa02b} Cs\'a, C., Kaloper, N., \& Terning, J., 2002, Phys. Lett. B, 535, 33
\bibitem[Dam et al.(2017)]{dam17} Dam, L.H., Heinesen, A., \& Wiltshire, D.L., 2017, MNRAS, 472m 835
\bibitem[Colin et al.(2019)]{col19} Colin, J., Mohayaee, R., Rameez, M., Sarkar, S., 2019, A\&A, 631, L13
\bibitem[Spallicci et al.(2021)]{spa21} Spallicci, A.D.A.M., Helay\"el-Neto, J.A., L\'opez-Corredoira, M., Capozziello, S, 2021, Eur. Phys. J. C, 81, 4
\bibitem[Spallicci et al.(2022)]{spa22} Spallicci, A.D.A.M., Sarracino, G., Capozzielo, S., 2022, Eur. Phys. J. Plus, 137, 253
\bibitem[Sarracino et al.(2022)]{sar22} Sarracino, G., Spallicci, A.D.A.M., Capoziello, S., 2022, Eur. Phys. J. Plus, 137, 1386

 \bibitem[van Putten(2015)]{van15a} van Putten, M.H.P.M., 2015a, Int. J. Mod. Phys. D, 3, 1550024

\bibitem[Chamblin \& Gibbons(1997)]{cha97} Chamblin, A., \& Gibbons, G.W., 1997, Phys. Rev. D, 55, 2177
\bibitem[Hod(1998)]{hod98} Hod, S., 1998, Phys. Rev. Lett., 81, 4293

\bibitem[Louko \& Marolf(1997)]{lou97} Louko, J., \& Marolf, Phys. Rev. D., 58, 024007 
\bibitem[Gibbons(2986)]{gib86} Gibbons, G.W., 1986, Nucl. Phys. B, 271, 497
\bibitem['t Hooft(2021)]{tho21} 't Hooft, G., Universe, 2021, 7, 298


\bibitem[Note(3)]{n3} 
$m$ is accompanied by an antiparticle $-m$ by conjugation $III=I^\dagger$ of wedges I and III in ${\cal M}$.
Equivalent to $2m$ moving forward in time, it accounts for the one-half period $\pi i$ in $\lambda$ in ${\cal R}$, 
versus $2\pi i$ in ${\cal M}$ by which ${\cal M}$ and ${\cal R}$ are distinct.
(On the bifurcation horizon $h$, this reduces particles to bosons.)

\bibitem[Note(4)]{n4} 
A particle in ${\cal R}$ carries an entanglement entropy $S=2\pi\varphi$ in Compton phase $\varphi=k_C\xi$  
defined by ita propagator \citep{van15a}. The associated probability ratio of emission and absorption across $h$ satisfies
$P_{e}=P_{a}e^{-S}$ in terms of the Boltzmann factor $e^{-S}$ \citep{har76,pad05}. 
The same identifies entropy changes when slowly dropping a test mass in 
a black hole \citep{bek73,bek81,ver11}. In ${\cal R}$, it establishes equality of $E=mc^2$, gravitational binding energy 
$U_h = ma\xi=mc^2$ and entropic work $W=\int_0^\xi T_UdS=T_US$. A photon at angular frequency $\omega$ and wavelength $\lambda$
similarly satisfies $\varphi = 2\pi\xi/\lambda$, giving $S=2\pi\varphi = \hbar\omega/ k_BT_U$ at the Davies-Unruh temperature 
$T_U$ upon squaring the Bogoliubov coefficient of the creation operator, $\left|\beta\right|^2 = 1/\left(e^S-1\right)$.
But see \cite{tho84,tho85} and (\ref{EQN_w}) in the present discussion.

\bibitem[van Putten(2017)]{van17} van Putten, M.H.P.M., 2017, ApJ, 837, 22

\bibitem[Bekenstein(1981)]{bek81} Bekenstein, J.D., 1981, Phys. Rev. D 23, 287
 \bibitem[Kokkotas \& Schmidt(1999)]{kok99} Kokkotas, K.D., Schmidt, B.G., 1999, Liv. Rev. Relat., 2, 2

\bibitem[Padmanabhan(2005)]{pad05} Padmanabhan, T., 2005, Phys. Rep., 406 49
\bibitem[Verlinde(2011)]{ver11} Verlinde, E., 2011, J. High Energy Phys., 4, 29
\bibitem[Hartle \& Hawking(1976)]{har76} Hartle, J.B., \& Hawking, S.W., 1976, Phys. Rev. D, 13, 2188

\bibitem[Bekenstein \& Mukhanov(1995)]{bek95} Bekenstein, J.D., \& Mukhanov, V.F., 1995, Phys. Lett. B, 360, 7

\bibitem[Spallicci \& van Putten(2016)]{spa16} Spallicci, A., \& van Putten, M.H.P.M., 2016, Int. J. Geom. Meth. Mod. Phys., 13, 1630014

\bibitem[van Putten(2015)]{van15b} van Putten, M.H.P.M., 2015b, arXiv:1506.08075
\bibitem[van Putten(2021)]{van21} van Putten, M.H.P.M., 2021, Phys. Lett. B, 823, 17, 136737
\bibitem[van Putten \& Della Valle(2023)]{van23} van Putten, M.H.P.M., \& Della Valle, M., 2023, A\&A, A36

\bibitem[Valdivia-Mera(2023)]{val23} Valdivia-Mera, G., 2023, arXiv:09869v2
 \bibitem[Chandrasekhar \& Detweiler(1975)]{cha75} Chandrasekhar, S., \& Detweiler, S., 1975, Proc. R. Soc. Lond. A, 344, 441
\bibitem[Schutz \& Will(1985)]{sch85} Schutz, B.F., \& Will, C.M., 1980, ApJ, 291, L33 %30
\bibitem[Kokkotas \& Schmidt(1999)]{kok99} Kokkotas, K.D., \& Schmidt, B.G., 1999, Liv. Rev. Rel., 1999-2
\bibitem[Leaver(1985)]{lea85} Leaver, E.W., 1985, Proc. R. Soc. Lond. Ser. A, 402, 285
\bibitem[Leaver(1986)]{lea86} Leaver, E.W., 1986, Phys. Rev. D, 34, 384
\bibitem[Nollert(1993)]{nol93} Nollert, H.-P., 1993, Phys. Rev. D, 47, 5253
\bibitem[Cardaso \& Gualtieri(2016)]{car16} Cardoso, V., \& Gualtieri, L., Class. Quant. Grav., 33, 174001
\bibitem[Berti et al.(2009]{ber09} Berti, E., Cardoso, V., \& Starinets, O., Class. Quant. Grav., 26, 163001

\bibitem[Echeverria(1989)]{ech89} Echeverria, F., Phys. Rev. D, 40, 3194
\bibitem[Dryer et al.(2004)]{dre04} Dreyer, O., Kelly, B.J., Krishnan. B., et al., 2004, Class. Quant. Grav., 21, 787
\bibitem[Berti et al.(2006)]{ber06} Berti, E., Cardoso, V., \& Will, C.M., 2006, Phys. Rev. D, 73, 064030
\bibitem[Berti et al.(2007)]{ber07} Berti, E., Cardoso, J., Cardoso, V., \& Cavaglia, M., 2007, Phys. Rev. D, 76, 104044

\bibitem[Ma et al.(2023)]{ma23} Ma, S., Sun, L., \& Chen, Y., 2023, Phys. Rev. Lett., 130, 14101

%\bibitem[Park et al.(2022)]{par22} Park, H.-J, Kim, S.-J., Kim, S., \& van Putten, M.H.P.M., 2022, ApJ, 938, 69

\end{thebibliography}

\end{document}